\documentstyle[preprint,prb,aps]{revtex}
\begin{document}
\draft
\title{Peak effect in CeRu$_2$:
history dependence and supercooling}
\author{S. B. Roy, P. Chaddah and Sujeet Chaudhary}
\address{Low Temperature Physics Laboratory,\\
Centre for Advanced Technology,\\ Indore 452013, India}
\maketitle
\date{\today}
\begin{abstract}
We present experimental results on single crystal CeRu$_2$ 
showing that the extent of 
history dependence of peak-effect depends on the 
path followed in the space of field (H) and temperature ( T ).
The (H,T) regime over which history effect is observed is larger if the 
vortex lattice is prepared by lowering T from above T$_C$ in 
constant H i.e. by field cooling. We compare this history 
effect with the very recently reported history dependence 
of peak-effect in detwinned single crystals of YBaCuO,      
highlighting the similarities and differences. We discuss 
the possibility of a first order vortex solid-solid transition  
in CeRu$_2$ within the realm of recent theoretical 
developments in the field of vortex matter.
\end{abstract}                          

\section{introduction}
Recent theoretical \cite{1,2,3,4} and experimental \cite{5,6,7} 
studies based on 
high T$_C$ superconductors (HTSC) suggest the existence of at least two
distinctly resolved solid phases of vortex matter which are distinguished 
from the high-temperature high-field vortex liquid. These two vortex solid
phases are referred to as low-field quasi-ordered solid or Bragg-glass and  
high-field disordered solid or vortex-glass (see Ref.8). 
The important question now
is, what is the the order of the thermodynamic phase transitions (if any)
between the various vortex phases? The Bragg-glass has long range order 
and it is expected to melt to vortex liquid at high temperature through a 
first order transition. Experimentally, the indication of a first order
transition usually comes via a hysteretic behaviour 
of various properties, not necessarily  thermodynamic ones. In HTSC samples
also initial suggestions of a first order melting transition came via distinct
hysteresis observed in transport property measurements \cite{9,10,11}. 
The confirmatory
tests of first order transition ofcourse 
involve the detection of discontinuous change in 
thermodynamic observables and the estimation of latent heat, and this has 
subsequently been achieved for vortex melting in 
HTSC materials through magnetization \cite{12,13} 
and calorimetric measurements \cite{14}. 
There also exists a less rigorous class of
experimental tests which involves the study of phase co-existence and 
supercooling across a first order transtion. This kind of experiment has 
also come out to be pretty informative for the first order melting transition
of the Bragg-glass in                             
Bi$_2$Sr$_2$CaCu$_2$O$_{8+\delta}$ (BSCCO) (Ref.15). 

With the establishment of the first order nature of the Bragg-glass to
vortex liquid transition line, the focus in the recent years 
has shifted to the Bragg-glass to vortex-glass transition. 
In various HTSC materials peak-effect(PE) or fish-tail is   
used to track this field induced transition from 
Bragg-glass to vortex-glass, and this transition is
obsereved to be a sharp transtion \cite{5,6,7}. However,
the exact nature of this transition
--whether it is a continuous or a first order transition--
is not established yet. Very recent magneto-optics 
studies on single crystal samples
of BSCCO claim the presence of phase-coexistence 
\cite{16,17} 
and supercooling \cite{17} across the Bragg-glass-vortex-glass 
phase transition. These in turn suggest the possibility of 
a first order phase transition.  
Newer theoretical developments \cite{18,19,20,21}
have also taken place in the field of vortex-matter 
physics  during last few years 
to understand the various field/disorder induced 
phenomena. This activity in vortex 
matter physics can even have deep correlations with a more 
general area namely disorder/pressure induced melting/amorphization in real 
solids \cite{22,23}.   

Very recently interesting history dependence of PE has been reported for 
naturally untwinned and detwinned single crystal samples of 
YBa$_2$Cu$_3$O$_y$(YBCO) with 6.908$\leq$y$\leq$6.999  
\cite{24,25,26}. 
We have earlier observed
very similar history dependence of PE in the low 
temperature (T$_C\approx$6.1K) C15-Laves phase 
superconductor CeRu$_2$ \cite{27,28,29}.
We believe that the $\it"minor$ $\it hysteresis$ $\it loop"$ technique 
used in our early study of
history dependence of PE in CeRu$_2$ \cite{27,28,29} 
and also in the similar recent studies on YBCO \cite{24,25,26}
belongs to that class of experiments which can investigate the phase
coexistence and supercooling across a first order phase transition.

Interesting Fermi surface topology and enhanced pauli paramagnetism 
of CeRu$_2$  have given rise to various interesting possibilities,   
starting from  an exotic non s-wave superconducting ground state \cite{30} 
to a field induced change in 
the microscopic superconducting order parameter associated  
with the onset of a Fulde-Ferrel-Larkin-Ovchinnikov (FFLO) 
state \cite{31,32,33}. Onset of FFLO state
can cause  softening of the vortex lattice 
and in turn enhanced pinning and PE \cite{34}. It is   
quite clear that if a field induced first order transtion 
from a BCS state to FFLO state actually takes place in any
system, a PE with definite history effects (typically associated with a
first order transition)  will be observed, and  
this has been motivating us, possibly as a red
herring, in our study of vortex matter in CeRu$_2$.
The question is can CeRu$_2$  
actually sustain a FFLO state ?
Our own macroscopic magnetization \cite{27,28,29} and transport-property 
measurements \cite{35} cannot provide any proof for (or against) 
the existance of FFLO state, and at this moment the global  
debate on this issue remain unsettled\cite{31,32,36,37,38}.

In the light of various recent developments described 
above regarding vortex matter, 
we believe that results obtained in CeRu$_2$ should be re-examined. 
In this paper we shall 
\begin{enumerate}
\item present newer results on the history dependence 
of PE in a single crystal sample of CeRu$_2$. Recently we have extended
the classical theory for supercooling across first order phase transitions
to the case when both field and temperature are control 
variables and have shown
how the observable region of metastability depends on the path followed in
the space of these two variables \cite{39}. We show that our
present experimental results are in
consonance with these theoretical predictions, and hence reinforce the 
idea
of a first order transition from one kind of vortex solid to another.
\item compare our results on CeRu$_2$ 
with the recently reported history dependence of PE 
in single crystal samples of YBCO \cite{24,26} and point out the 
similarity and difference   
between these two disparate class of materials. The relevance of the results
on the phase coexistence and supercooling in BSCCO \cite{16,17}
will also be discussed.
\item discuss a possible origin of a first order transition 
between two kinds of vortex solid in CeRu$_2$
within the realm of the recent theoretical developments \cite{18,19,20,21}. 
\end{enumerate}
Preliminary results of the present work have 
been presented in the recently held LT-22 conference in Helsinki \cite{40}.

\section{experimental}
In contrast to the polycrystalline samples of CeRu$_2$ used in  
our earlier studies \cite{27,28,29} showing history effects associated 
with PE, in the present study we use a single crystal sample 
of CeRu$_2$ (T$_C \approx$6.1K). The details of the 
preparation and characterization of this sample can be found in Ref.41. 

Magnetization measurements were performed using a 
commercial SQUID magnetometer (Quantum Design MPMS5). We    
have used a 2 cm scan length in the 'fixed-range' mode to   
minimize the sample movement in the inhomogeneous field of 
the superconducting magnet. In the 'auto-range' mode the    
sample goes through multiple movements while the system 
software searches for the most sensitive gain useful for the 
signal level detected. We carried out a separate preliminary 
run using the auto-range mode to identify the appropriate 
gain for the given experimental conditions and then 
performed a final run in the 'fixed-range' mode. In the case 
of  2cm scan-length, the field inhomogeneity in an applied 
field of 20 kOe is $\approx$2 Oe. We have concluded earlier \cite{42}
that in an isothermal field scan, as long as the field for  
full penetration at a particular field value is             
substantially larger than the field inhomogeneity           
during the sample measurement, the error in the results in 
the particular type of measurements reported here will be 
negligible. Inspite of all these cross-checks, 
in the light of general doubts \cite{43,44} 
concerning the measurement procedure using commercial SQUID magnetometers, it 
has become important to reproduce the observed history effects 
using other techniques which minimize the sample 
movement. In a recent work \cite{45} we have 
shown the existence of the history dependence of PE 
in the isothermal field variation of magnetization 
in polycrystalline samples of CeRu$_2$ using
an axial-VSM (Oxford Instruments). 
In the present work we have used a transverse-VSM (Oxford Instruments) to 
get supporting results. In this transverse VSM 
the sample is placed in the centre of 
the pick-up coil and the superconducting magnet assembly, where 
the magnetic field inhomogeneity is $\approx$0.01\%   
over 1 cm diameter spherical volume (DSV). 

In contrast to the axial-VSM, 
the direction of 
sample vibration in the transverse-VSM is perpendicular to the applied field.
No significant change is observed in the results by varying the 
sample vibration amplitude between 0.5 and 1.5mm; this rules out any distinct
role of magnetic field inhomogeneity. 
In any case the field inhomogeneity encountered here is obviously much  
smaller than that encountered even in the 2cm-scan of SQUID 
magnetometer.

In magnetization hysteresis measurements, we draw an isothermal magnetization
(M) versus field (H) curve by cycling H between $\pm H_{C2}$ at various 
temperatures (T)  below T$_C$. These T of interest are reached 
by cooling through T$_C$ in absence of any applied field H i.e. in zero-field-
cooled (ZFC) mode. From various H-points on this isothermal M-H curve (or 
envelope curve), a $\it minor$ $\it hysteresis$ $\it loop$ (MHL) 
can be be drawn either
by decreasing H from the ascending field branch of the M-H curve 
(or lower envelope curve) or
by increasing H from the descending field branch (or upper envelope curve).
We designate these MHLs as (MHL)$_{ZFC}$. One can also initiate MHLs from
various H-points obtained by field cooling (FC) through T$_C$. In such a case
the M-value at the starting H-point normally lies 
inbetween the upper and lower envelope curve 
(see Ref.27 and references cited therein). 
 An MHL can be initiated from such FC H-points either by 
increasing or by decreasing H. We designate these MHLs as (MHL)$_{FC}$. Within
the realm of a single-component Bean's crtical state model, all these MHLs
(both ZFC and FC) are expected to saturate by reaching the 
enevelope curve \cite{46}. Such a behaviour has actually been 
observed experimentally  in various type-II superconductors . The same
is also observed  in all H-regimes, except a finite field regime encompassing 
at least a part of the PE regime, in various samples of CeRu$_2$ \cite{27}
including the present single
crystal sample. In this latter regime the MHLs do
not behave in accordance with the critical state models, and we have 
used this anomalous behaviour of MHLs to study the interesting history
dependence of PE in CeRu$_2$ \cite{27,28,29}. 
Very similar technique involving the 
(MHL)$_{ZFC}$ has very recently been used to study the history dependence of
PE in single crystals of YBCO \cite{24,25,26}. 

Apart from tracking 
the history dependence of PE the study of MHLs can provide with 
at least two more useful information in our present kind of study. First, it
provides a way of estimating the field for full penetration at a particular
H of interest \cite{27,42}. 
The field inhomogeneity $\delta$H of the magnet causes the sample to 
effectively follow an MHL during measurements 
with a SQUID magnetometer \cite{47}. Since $\delta$H rises with scan length,
MHLs allow one to estimate the error $\delta$M in measurements made with
various scan lengths. One can then choose a scan length such that $\delta$M is
much smaller than the magnetization hysteresis, or, equivalently $\delta$H is
much smaller than the field for full penetration. 
Second, a fair amount of information regarding the
influence of surface barrier can be obtained from the nature of the approach
(linear or non-linear) of these MHLs to the envelope curve \cite{48}. These 
two information are useful even where the M-H curves are
in accordance with the critical state models \cite{46}. Thus the technique of
MHLs has universal applicability in determining (i) the effect of field 
inhomogeneity of the magnet on a measurement (ii) the importance of surface
barrier in the magnetization study.

While presenting newer results, the present study on a single 
crystal sample of  CeRu$_2$ using both  SQUID magnetometer and VSM, provides
also a cross-check on our earlier results \cite{27,28,29} 
obtained on polycrystalline samples
mainly using a SQUID magnetometer. This will put the experimental situation
regarding the PE in CeRu$_2$ on a more firm ground.

\section{Results and discussion}
Fig. 1 shows M-H  plots of the 
CeRu$_2$ single crystal at T=4.5K, obtained using both       
SQUID magnetometer and VSM. These M-H curves are obtained by   
isothermally cycling H between $\pm$25kOe. Since           
H$_{C2}(4.5K)\approx$21.5kOe is less than 25 kOe, these      
provide the envelope hysteresis curve within which all 
the MHLs should be contained. This envelope M-H curve 
shows two distinct irreversible regimes separated by an 
almost reversible regime (see Fig. 1). While this           
intermediate regime appears quite reversible in the SQUID   
measurement (see inset of Fig. 1(a)), perceptible 
irreversibility is observed in the VSM measurement (see inset of Fig. 1(b)). 
This field-induced enhanced magnetization-irreversibility 
in the high field regime is the so 
called peak-effect (PE) and this is the subject of main 
interest in the present work. We note that in Fig.1 the 
onset field H$^*_a$ of the PE in the ascending field cycle is distinctly 
different from the field H$^*_d$ where the PE ceases to 
exist in the descending field cycle.
Note that H$^*_a$ and H$^*_d$ obtained from the SQUID and VSM measurements
are a bit different. We attribute this quantitative difference to 
(i)the difference in the magnetic field inhomogeneity encountered during
SQUID and VSM measurements, (ii)possible minor difference in temperature in
two different machines. Also the magnitude of measured M is smaller in the VSM
measurement which we attribute to different sample orientation and the 
associated demagnetization factor. However, it should be noted
that the actual measurement time involved in a VSM measurement,
where the field is swept with a constant rate (100 Oe/s in the
present case) is faster than in a SQUID measurement where the
field is stabilized with a pause time (10 sec in the present
case) before each measurement. In systems with finite
magnetization relaxation, faster measurement with VSM would
yield larger magnetization value. This is actually observed in
the field regime just below the PE where magnetization
hysteresis obtained with VSM measurements is perceptibly higher
[see Fig.1(a) and Fig.1(b)].

We shall now present results obtained in the form of MHLs 
measured at closely spaced field intervals, after preparing 
the vortex lattice within the following experimental        
protocols :
\begin{enumerate}
\item Zero field cool (ZFC) the sample to the temperature    
of measurement, switch on a field less than -H$_{C2}$, and  
then increase the field isothermally to reach various points 
on the lower envelope curve. (MHL)$_{ZFC}$'s are drawn by reducing the 
field isothermally.
\item After the above step, increase the field to a value   
greater than H$_{C2}$ and then reduce the field 
to reach the various points on the upper envelope curve,    
while maintaining the isothermal condition. (MHL)$_{ZFC}$'s are 
drawn by increasing the field isothermally.
\item Field cool (FC) the sample in various 
(positive) fields from a temperature substantially above    
T$_C$. After stabilizing the temperature of interest 
the (MHL)$_{FC}$'s can be drawn both by increasing and decreasing  
the field isothermally.
\end{enumerate}

The MHLs at the onset of the PE regime obtained within the 
above experimental protocols 
do not conform with the critical-state models;they do not   
show the expected merger with the envelope curve (see Fig.2). 
While the (MHL)$_{ZFC}$'s initiated from
the lower envelope curve at H=18.25, 18.75 and 19 kOe saturate without 
touching the upper envelope curve (see Fig. 2(a)), the (MHL)$_{ZFC}$'s 
initiated from the upper
envelope curve at H=18.4 kOe overshoots the lower envelope curve 
before reaching saturation (see Fig. 2(b)).
The (MHL)$_{FC}$'s obtained following the FC path also overshoot the 
envelope curve (see Fig. 2(c)). For the sake of clarity and conciseness we
show only few representative MHLs.

We have earlier reported \cite{27,28} anomalous behaviour of (MHL)$_{ZFC}$'s 
obtained within the protocol no.1 in polycrystalline samples of 
CeRu$_2$. Tenya et al \cite{49} have reported anomalous (MHL)$_{ZFC}$s under 
protocol no.2. Ravikumar et al (Ref.50) and ourselves     
(Ref.29(b)) reported anomalous (MHL)$_{FC}$'s under protocol no.3 in single 
crystal and polycrystalline samples of CeRu$_2$ respectively. 
Actually the anomalous         
character of the (MHL)$_{ZFC}$ initiated from the upper envelope curve 
was visible in a   
relatively less prominent manner in our earlier studies of 
polycrystalline samples as well (see Fig.3 of Ref.29 (a)). 
However, from our standard cross-checks we were not sure     
whether the observed result was beyond error bar, hence did 
not emphasise much on that. (On the other hand we were aware of the
metastable character of those (MHL)$_{ZFC}$'s drawn from the upper envelope
curve, since they readily shattered on field cycling 
(see Fig. 3 of Ref. 29(a).) 
In the present work for the first time all the three different kinds 
of (MHL)s are shown on the same sample of CeRu$_2$. 

We shall now reproduce all these anomalous aspects of MHLs in the 
vicinity of PE  using a transverse-VSM. 
Fig.3 shows the anomalous nature of the two (MHL)$_{ZFC}$'s drawn
from the lower envelope curve at H=19.25 kOe and 20 kOe, and 
two (MHL)$_{FC}$'s 
drawn by reducing the field at H=17.25 and 19 kOe. 
In this transverse-VSM it is relatively difficult
to stabilize the temperature for T $<$10K and actual temperature 
can vary by $\pm$0.025K between different runs and some time even
during a single complete run. This leads to a slight 
scatter in the data, but from multiple runs of the same experiment we 
ensure that the observed anomalous features are certainly well beyond the 
error bars. For the two kinds of MHLs described above we need to obtain the
envelope M-H curve in a separate experimental cycle. However, the anomalous
nature of the (MHL)$_{ZFC}$'s drawn from the upper 
envelope curve can be highlighted
from a single experimental cycle. For this purpose we increase the field
isothermally to field values well above H$_{C2}$ (thus drawing the lower
envelope curve) and then reach the various field points of interest on the 
upper envelope curve by isothermal reduction of the field. From such field
points we draw (MHL)$_{ZFC}$'s by reversing the field sweep direction. 
In Fig.4 we show the MHLs initiated from the upper envelope curve at 
H=18.75 and 19.25 kOe  and they distinctly overshoot 
the lower envelope curve before saturation. 
These representative MHLS provide support to the results 
obtained earlier with the SQUID magnetometer.
Various envelope curves and MHLs presented in Fig.3 and
4 are obtained with field sweep rate 100 Oe/sec. 
We have also checked the qualitative aspects of our results by varying the
field sweep rate between 20 and 200 Oe/sec.

Interesting history effects
associated with the PE have been reported recently 
in detwinned and naturally untwinned single crystals
of YBCO \cite{24,25,26}. These history effects in YBCO are exactly 
in the same form of anomalous isothermal (MHL)$_{ZFC}$'s drawn from the upper 
and lower envelope M-H curve of CeRu$_2$ as shown in 
Fig. 2(a), 2(b), 3(a) and 4. In these reports, however,
there is no mention of any measurement involving    
MHLs in a field cooled vortex state.

We have earlier associated these anomalous features 
in polycrystalline samples of CeRu$_2$ with a field induced first
order transition to a superconducting mixed state with 
enhanced pinning properties \cite{27,28,29}.
While the multivaluedness in the 
saturation of the (MHL)$_{ZFC}$ in the ascending field cycle
was attributed to the nucleation and growth of the high field phase, the 
overshooting of the envelope M-H curve by the (MHL)$_{ZFC}$'s 
in the descending field cycle and by 
the (MHL)$_{FC}$, was thought to be a result 
of supercooling of the high field phase across the
proposed first order transition \cite{27,28,29}.
The present results on a good quality single crystal of CeRu$_2$ reinforce
this picture.

We shall now provide newer evidence to support the conjecture 
of a first order phase transition in CeRu$_2$. Extending the 
classical theory for supercooling across first order phase transitions
to the case when both field and temperature are control variables, we have
shown theoretically that the observable region of metastability depends on
the path followed in this space of two variables, with variation of
field providing a source of fluctuations in the supercooled
state \cite{39}. We have predicted that a
disordered phase can be supercooled upto the limit of metastability T$^*$(H)
only if T is lowered in constant H. If the T$_C$(H) line is crossed by
lowering H at constant T, then supercooling will terminate at T$_0$(H) which
lies above the T$^*$(H) line \cite{39}. If T$_C$ falls with rising field, 
then (T$_0$(H)-T$^*$(H)) rises with rising field \cite{39}. 
As narrated below both these predictions are
experimentally found to be true in CeRu$_2$.    
We have found that
the anomalous features in (MHL)$_{FC}$ in CeRu$_2$ continue to 
exist in a (H,T) regime which is well below the PE regime. In this regime 
the (MHL)$_{ZFC}$'s show normal behaviour as expected within the   
critical state models. This clearly suggests that the
FC vortex state can be supercooled more than
the isothermal ZFC state.  Collating the H values where the 
various MHLs first show the anomalous behaviour at various 
T in our SQUID magnetometer measurements, 
in Fig.5 we present a (H,T) phase diagram. The distinct     
identity of the H$^*_a$(T) line (which indicates the onset 
of the PE regime in the isothermal ascending field cycle) 
and the H$^*_d$(T) line (at which the PE regime terminates  
in the isothermal descending field cycle) was infact earlier 
taken as an indication of a first order transition \cite{31,32}. 
H$^*_d$(T) and H$^*_{FC}$(T) lines are akin to the T$_0$(H) 
and T$^*(H)$ lines respectively in our theoretical study \cite{39}.
Experimentally H$^*_{FC}$ at a particular temperature is defined as 
the H value down to which the anomalous behaviour in the (MHL)$_{FC}$'s
is observed. Similarly below H$^*_d$(T) (defined earlier) the (MHL)$_{ZFC}$'s
drawn from the upper envelope curve show normal behaviour namely, they
merge with the lower envelope curve without any overshooting. 
As shown in Fig.5  the H$^*_{FC}$(T) line is lying distinctly below 
the  H$^*_d$(T) line and (H$^*_d$(T) - H$^*_{FC}$(T)) increases 
with the increase in H. 
Although anomalous behaviour of (MHL)$_{ZFC}$'s and (MHL)$_{FC}$'s in
CeRu$_2$ has been highlighted by various groups during last 
three years\cite{27,28,29,49,50}, 
the path dependence of this anomalous behaviour in (H,T)
space is definitely new and has not been reported so far.    
These new results, which are in accordance with our theoretical 
prediction \cite{39}, will provide further support for the existence of a 
first order transition in the vortex matter phase diagram of CeRu$_2$. 

The observed history effects in YBCO are interpreted in terms of a 
transition from a low field elastic vortex lattice to a high 
field plastic vortex lattice \cite{24,25,26}. 
We have previously discussed this  
mechanism in the context of PE in polycrystalline samples of CeRu$_2$, 
and argued that a first order 
transition probably has an edge over this mechanism in explaining the 
history dependent phenomena associated with PE (see Ref.29(b)). Since 
then our view point is reinforced by the transport study    
of the PE on a single crystal sample of CeRu$_2$ \cite{35}.  
We have shown that the FC state at the vicinity of the PE regime is 
metastable in nature and it can be shattered easily with a 
small field cycling (see Ref. 29(b) and 35). 
This observation can be relatively 
easily explained in terms of the metastable nature of a 
supercooled state and its sensitivity to any environmental 
fluctuation \cite{39}.
In a very recent study of magnetization in untwinned single
crystal of YBCO using micro Hall probe, history effects in FC
measurements have now been reported \cite{51}. Relatively
complicated (H,T) phase diagram of YBCO \cite{52} needs a
contrived path to ensure proper FC measurements \cite{51} and
this probably may be the reason why FC measurements were not
reported in earlier studies \cite{24,25,26}. In contrast to the
suggestion that the observed history effects are properties of
the high field plastic vortex lattice \cite{24,25,26}, it is
asserted that both the low field and the high field vortex
lattice are robust in nature, and the history effects and
metastability are associated with the transition regime from the
low field to high field phase\cite{51}. This recent observation makes the
possibility of a first order transition in YBCO much stronger,
and hence the similarity with CeRu$_2$.

Contrary to our earlier suggestion based on 
bulk magnetization study \cite{29}, 
a very recent magneto optical study in good single 
crystals of BSCCO has claimed that the solid-solid
transition in the vortex matter is indeed a first order transtion accompanied
by supercooling \cite{17}. If this claim turns out to be true,
it is possible that the supercooled (H,T) regime in
BSCCO is quite narrow and/or fragile. So in our magnetization measurement
we have either missed the supercooled 
regime and/or the fluctuation induced during the
measurement procedure might have shattered the supercooled phase.    

At this juncture we must point out some important 
difference in the history effects associated 
with PE in CeRu$_2$ and YBCO. While 
the history effects associated with PE were observed only in 
naturally untwinned and detwinned single crystals of YBCO and
vanish quite readily with the change in oxygen 
stoichiometry \cite{26}, the same effects
are quite robust in CeRu$_2$ and are observed with all the 
characteristic  features
in good quality single crystal, polycrystal, off-stoichiometric polycrystal
and Nd-doped polycrystal samples of CeRu$_2$
\cite{27,28,29,37,52}. These results suggest that, in contrast
with CeRu$_2$ the characteristic features associated with the
vortex solid-solid transition in YBCO are quite sensitive to the
defect. Also the negative dynamic creep at the onset of PE regime of YBCO 
(as reported recently (Ref.25)), is not observed with identical 
measurements on CeRu$_2$ \cite{54}. On the other hand, while the PE ceases to
exist above a temperature T$^* \approx$0.92 T$_C$ \cite{31,32} 
in all kinds of samples of CeRu$_2$, there is no report (to our knowledge) 
so far of a vortex-solid to vortex-liquid melting transition in CeRu$_2$. 
  
We shall now discuss some recent  
developments in the field of vortex matter physics \cite{18,19,20,21}
since they appear relevant to our experimental results in CeRu$_2$. 
At low tempeartures and at high fields 
disorder dominates in vortex matter and
topological defects proliferate, resulting in highly disordered solid .
We shall now explore the possibility of a disorder induced 
first order transition in the vortex matter which can lead to a 
softening of the vortex lattice. PE will be used as an observable effect 
of such lattice softening. (This is in contrast with the 
field induced FFLO state where the lattice softening is due 
to a micorscopic change in the superconducting order 
parameter \cite{34}.)
The point defects in the underlying crystalline lattice can cause transverse
wandering of the vortex lines  and this frozen-in wanderings can destroy
the long-range order of the vortex lattice (Ref.4). This is analogous to the
action of the thermal noise and can generate topological defects in the vortex
lattice. The role of topological defects in the             
superconducting mixed state of type-II superconductors has  
been a subject of interest over the years \cite{55,56}
and has come under closer scrutiny recently in the 
context of vortex-solid \cite{18,19,20,21}. 
Frey, Nelson and Fisher \cite{18} suggested that 
in the low temperature-high field regime of vortex matter,
topological defects in the form of vacancies
and interstitials can start proliferating leading to an 
intermediate supersolid state. The vortex supersolid is characterized
by the coexistence of crystalline order and a finite equilibrium
density of vacancy and interstitial defects \cite{18,20}. 
The vortex supersolid then can transform continuously 
into a vortex liquid state \cite{20,57}.
Vortex liquid state is characterized by unbound dislocation loops as well as
finite density of vacancy and interstitials \cite{20}. The exact nature of
the transition from defect free vortex-solid to 
supersolid transition is not quite clear and
possibility exist for  both a continuous and a first order 
transition \cite{18}. Caruzzo and Yu \cite{19} have also 
considered the possibility  of a first order transition to a supersoftened
solid induced by interstitial and vacancy line defects in vortex lattice. 
Although Caruzzo and Yu \cite{19} mainly discuss the cases of phase transitions 
as a function of temperature, their theoretical approach actually 
is based on earlier works on disorder induced softening and 
first order transition in real solids \cite{58}. 
A supersoftened solid can be quite relevant in our present discussion
on the field/disorder induced transitions in 
vortex soilds of various superconductors.

It will not be totally out of place to mention here that 
defect induced melting and solid state amorphization is a 
distinct possibility in real solids and has remained a subject of    
continued interest \cite{22,23}. A first order transition in such 
cases can be characterized by a discontinuous increase in  point defects. 
 
\section{Conclusion}
Based on our results concerning history dependence of PE,  
we suggest the existence of a first 
order phase transition from one kind of vortex solid to   
another in the vortex matter phase diagram of CeRu$_2$. The high-field 
phase can be supercooled by reducing either of the two control variables
viz H and T. The extent of supercooling observed depends on the path followed
in this space of two variables. In the light of very recent observation of
history dependence of PE in detwinned sample of YBCO both in the
ZFC \cite{24,25,26,51} and FC measurements \cite{51}, it will be now of 
interest to check experimentally for a similar (T,H) path dependence and
possibility of a first order solid-solid phase transition in the vortex phase
diagram of YBCO.  

\section{acknowledgement}
We acknowledge Mohammed, S. Hebert, G. Perkins, L. F. Cohen and A. D. Caplin
for various help in the experiments involving the transverse-VSM and many
useful discussion. We also acknowledge Y. Radzyner, D. Giller, A. Shaulov and
Y. Yeshurun for useful discussion. 
We thank  Dr. A. D. Huxley for providing us with the 
single crystal sample of CeRu$_2$ used in the present study.

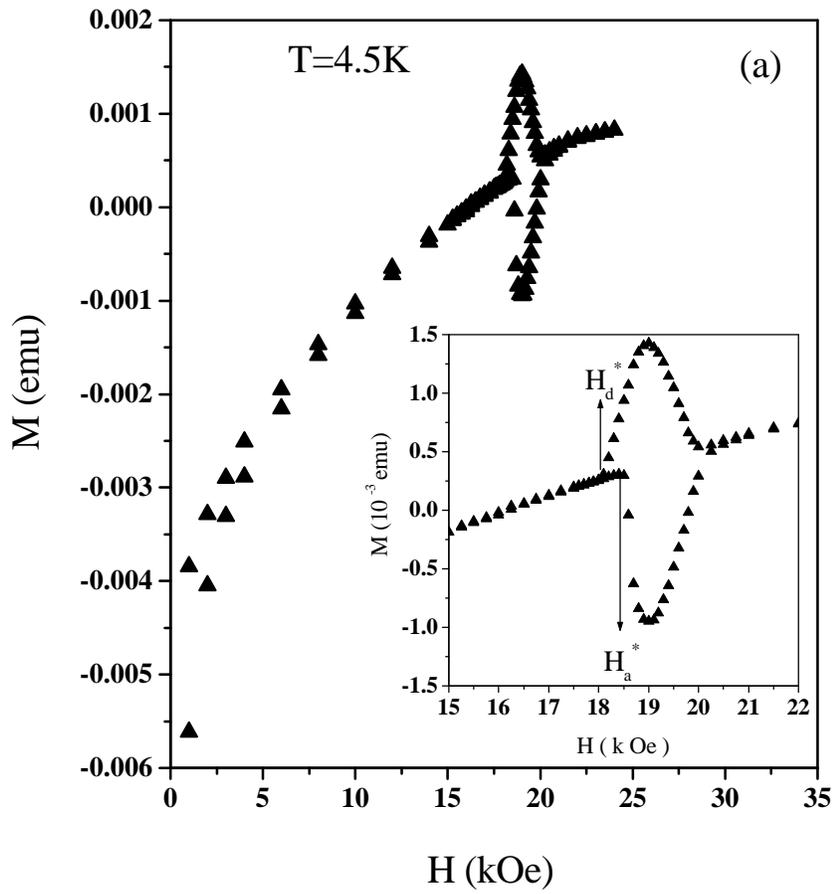
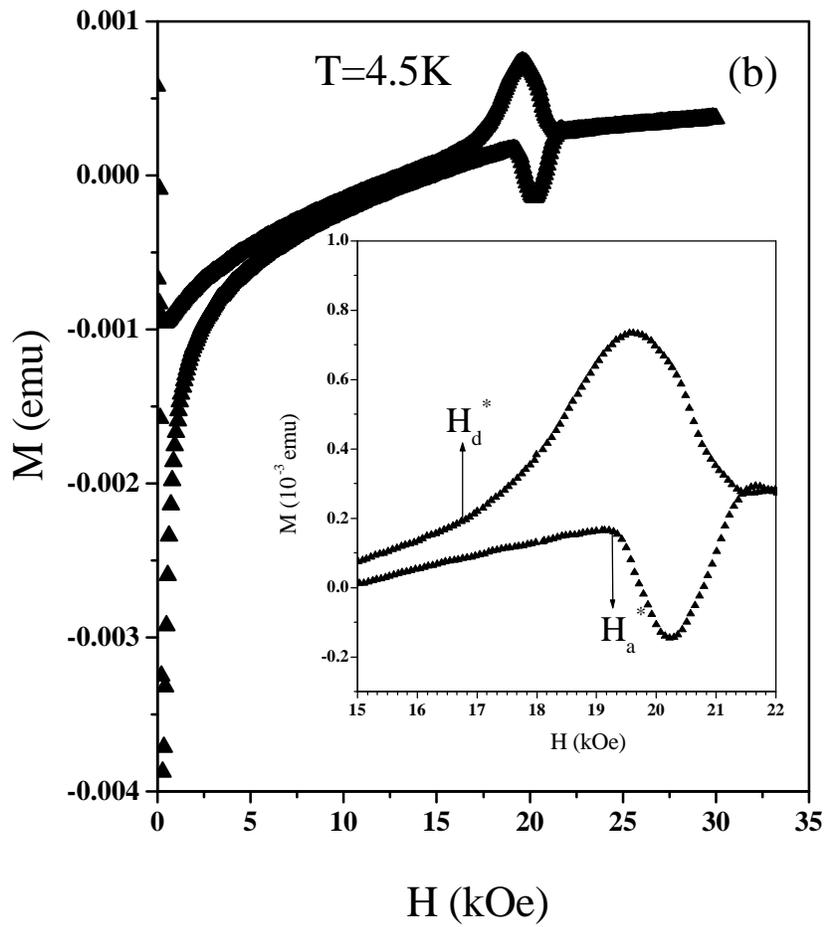
\begin{figure}
\caption{Magnetization (M) vs field (H) plot for CeRu$_2$ 
at 4.5K obtained with (a) SQUID magnetometer (b) vibrating  
sample magnetometer. See text for details. The insets 
highlight the peak-effect regime.}
\end{figure}
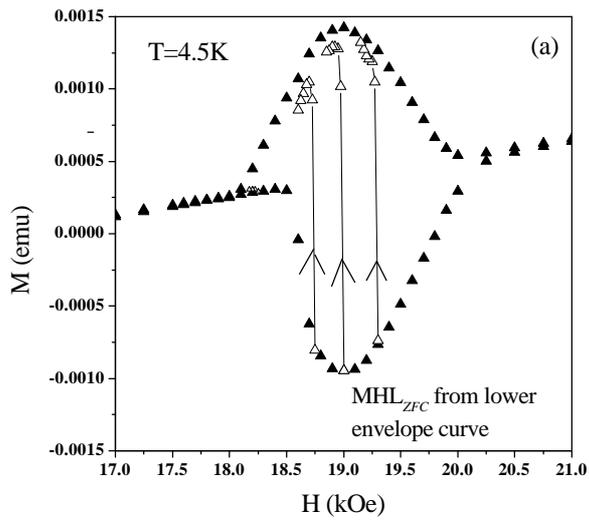
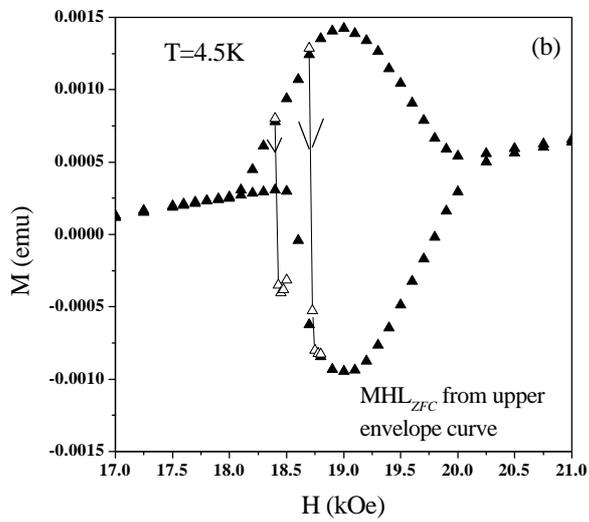
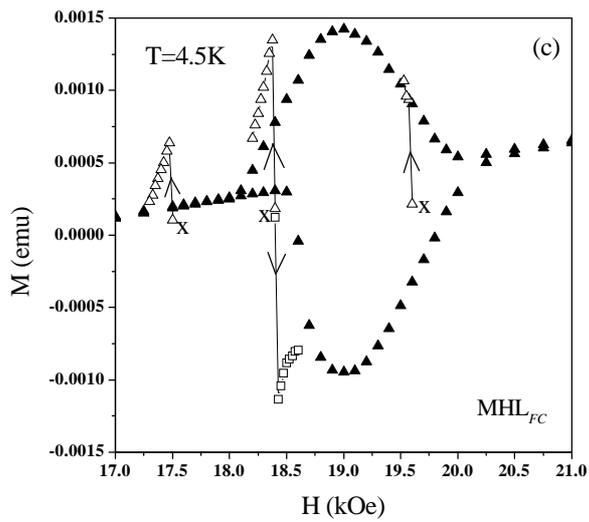
\begin{figure}
\caption{Forward legs of various minor hysteresis loops     
(MHL) at 4.5K obtained with SQUID magnetometer. (a) 
Open triangles denote (MHL)$_{ZFC}$ initiated from 
the lower envelope curve at H= 18.25, 18.75 and 19kOe; they saturate without 
touching the 
upper envelope curve. (MHL)$_{ZFC}$ initated from 19.6 kOe saturates on
touching the upper envelope curve. (b)Open triangles 
denote (MHL)$_{ZFC}$ initiated 
from the upper envelope curve at H=18.4 and 18.75 kOe. The (MHL)$_{ZFC}$
initiated at H=18.4 kOe overshoots
the lower envelope curve. (c)Open triangles denote (MHL)$_{FC}$        
initiated on decreasing field form H = 17.5, 18.4 and 19.6 kOe after field    
cooling. The starting points of these MHLS are marked with X.
While the first two MHLs overshoot the upper envelope curve, the one initiated
at H=19.6 kOe behaves normally, namely it saturates on touching the 
upper envelope curve. We also show
one (MHL)$_{FC}$ (represnted by open squares) 
initiated from H= 18.4 kOe by increasing H. See text for 
details. Filled triangles represent the envelope curve.} 
\end{figure} 
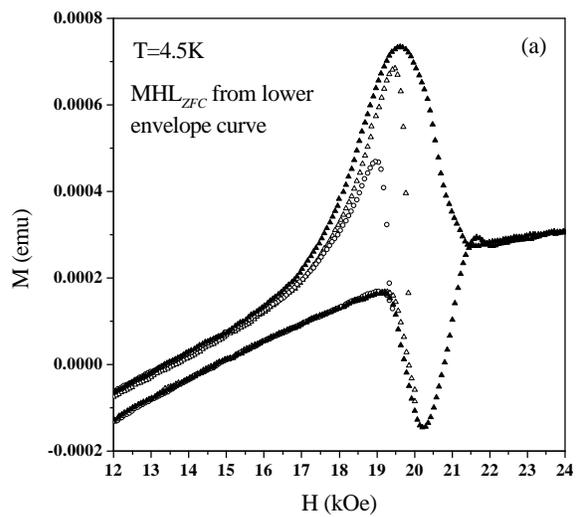
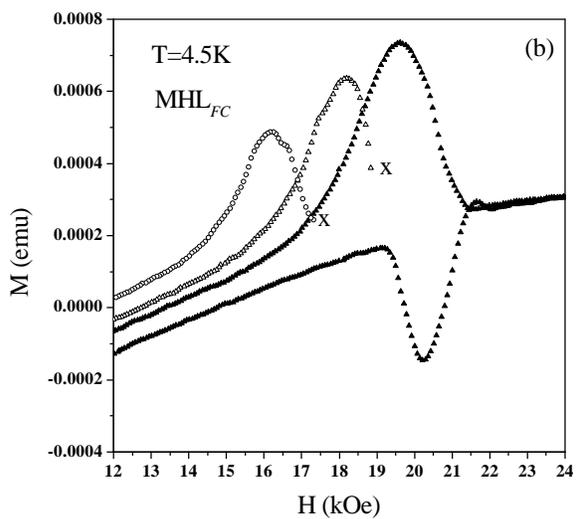
\begin{figure}
\caption{Forward legs of various minor hysteresis loops     
(MHL) at 4.5K obtained with vibrating sample magnetometer. 
(a)(MHL)$_{ZFC}$ initiated from the lower       
envelope curve at H=19.25 (open circle) and 20 kOe (open triangle);
as in the case of SQUID           
magnetometer measurements the MHLs saturate clearly below   
the upper envelope curve and meet the envelope curve after  
crossing the peak-effect regime. (b)(MHL)$_{FC}$        
initiated on decreasing field form H =17.25 (open circle) 
and 19 kOe (open triangle) after field    
cooling; they overshoot the envelope curve. 
X marks the starting  points of the (MHL)$_{FC}$'s.
Filled triangles represent the envelope curve.} 
\end{figure}
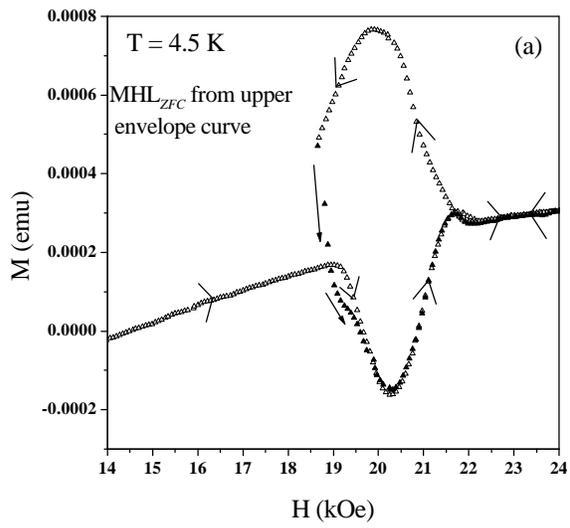
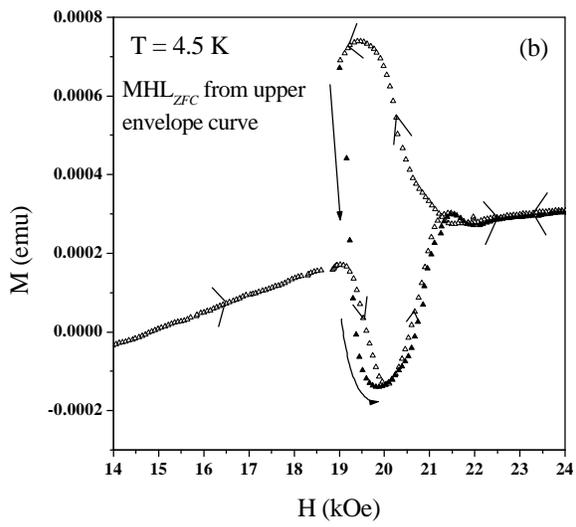
\begin{figure}
\caption{Minor hysteresis loops (MHL)$_{ZFC}$ drawn at 4.5K
(obtained with VSM) after increasing H isothermally from 0 to well above 
H$_{C2}$ (thus drawing the lower envelope curve) and then    
reaching the various field points of interest on the upper  
envelope curve by isothermal reduction of H. The MHLs are   
drawn be reversing the field sweep direction at (a) 
H=18.75 kOe. and (b) H=19.25kOe. The envelope curve is represented by
open triangle and the (MHL)$_{ZFC}$ by filled triangle.
The MHLs distinctly overshoot the lower envelope curve.}
\end{figure}
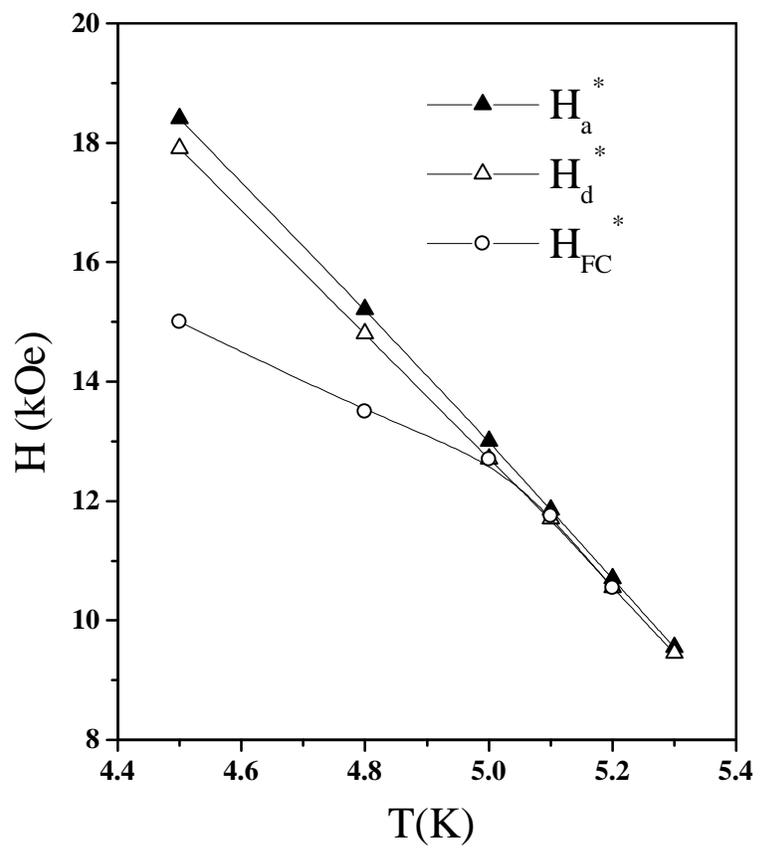
\begin{figure}
\caption{Experimentally obtained H-T phase diagram of CeRu$_2$, highlighting 
the history dpendence in the peak-effect regime. See text for details.}
\end{figure}
\end{document}